\documentclass[11pt]{amsart}

 \usepackage{graphicx}

\usepackage{amssymb}
\usepackage{eurosym}
\usepackage{amsthm}
\usepackage[left=1in, right=1in, top=1in, bottom=1in]{geometry}

\usepackage{lineno}

\usepackage{hyperref}

\usepackage{amssymb}
\usepackage{amsthm}
\usepackage{color}
\usepackage[utf8]{inputenc}
\usepackage{amsmath}
\usepackage{amsfonts}
\usepackage{color}
\usepackage{enumerate}
\usepackage{lineno}
\usepackage{graphicx}
\usepackage{multicol}
\usepackage{lipsum}
\usepackage{graphicx}
\usepackage{verbatim}
\usepackage{float}
\usepackage{booktabs}
\usepackage[usenames,dvipsnames]{xcolor}
\usepackage{amsmath, multicol, tikz, tikz-3dplot, pgfplots, amssymb, latexsym, amsmath, graphicx, verbatim, epsfig, url, mdwlist, setspace, placeins, xcolor, fancyvrb, amsthm, wrapfig, enumitem, multicol, graphicx}
\usepackage{latexsym,amssymb,amsmath,amsfonts,bm}
\usepackage{amssymb}
\usepackage{amsmath}
\usepackage{graphicx}
\usepackage{dcolumn}
\usepackage{hyperref}
\usepackage[utf8]{inputenc}
\usepackage[english]{babel}
\usepackage{amsthm}
\usepackage{latexsym}
\usepackage{xcolor}
\usepackage{flexisym}
\usepackage{ulem}

\usepackage{float}

\hypersetup{
	colorlinks,
	linkcolor={red!50!black},
	citecolor={violet!50!black},
	urlcolor={blue!50!black}
	}
	
\pgfplotsset{compat=1.18} 

\theoremstyle{definition}

\theoremstyle{remark}

\newcommand{\amanda}[1]{{\color{red} \sf $\int$ Amanda: [#1]}}
\newcommand{\iris}[1]{{\color{CarnationPink} \sf $\pi$ Iris: [#1]}}
\newcommand{\sam}[1]{{\color{green} \sf $\sigma$ Sam: [#1]}}
\newcommand{\audrey}[1]{{\color{cyan} \sf Audrey: [#1]}}

\newcommand{\josh}[1]{{\color{violet} \sf $\aleph$ Josh: [#1]}}

\definecolor{red}{RGB}{209,97,99}
\definecolor{purple}{RGB}{161,97,209}
\definecolor{blue}{RGB}{97,141,209}
\definecolor{blue2}{RGB}{97,174,209}

\begin{document}


\title{Predictive Modeling of Lower-Level English Club Soccer Using Crowd-Sourced Player Valuations}

\author[Josh Brown]{Josh Brown}
\address[Josh Brown]{Ursinus College}
\email{jobrown@ursinus.edu}

\author[Yutong Bu]{Yutong Bu}
\address[Yutong Bu]{Emory University}
\email{ybu4@emory.edu}

\author[Zachary Cheesman]{Zachary Cheesman}
\address[Zachary Cheesman]{Bowdoin College}
\email{zcheesman@bowdoin.edu}

\author[Benjamin Orman]{Benjamin Orman}
\address[Benjamin Orman]{Grinnell College}
\email{ormanben@grinnell.edu}

\author[Iris Horng]{Iris Horng}
\address[Iris Horng]{University of Pennsylvania}
\email{ihorng@wharton.upenn.edu}

\author[Samuel Thomas]{Samuel Thomas}
\address[Samuel Thomas]{Brown University}
\email{samuel_thomas@brown.edu}

\author[Amanda Harsy]{Amanda Harsy}
\address[Amanda Harsy]{Lewis University}
\email{harsyram@lewisu.edu}

\author[Adam Schultze]{Adam Schultze}
\address[Adam Schultze]{Lewis University}
\email{aschultze@lewisu.edu}

\begin{abstract}
In this research, we examine the capabilities of different mathematical models to accurately predict various levels of the English football pyramid. Existing work has largely focused on top-level play in European leagues; however, our work analyzes teams throughout the entire English Football League system. We modeled team performance using weighted Colley and Massey ranking methods which incorporate player valuations from the widely-used website Transfermarkt to predict game outcomes. Our initial analysis found that lower leagues are more difficult to forecast in general. Yet, after removing dominant outlier teams from the analysis, we found that top leagues were just as difficult to predict as lower leagues. We also extended our findings using data from multiple German and Scottish leagues. Finally, we discuss reasons to doubt attributing Transfermarkt's predictive value to wisdom of the crowd.
\end{abstract}



\maketitle



\section{Introduction}

Since its creation in 1888, the English Football League (EFL) has developed into a robust pyramid of interconnected football competitions \cite{Efl}. This structure includes many leagues at different tiers of play. The Premier League sits atop this pyramid. Directly below it is the modern EFL, consisting of the English Championship (tier 2), League One (tier 3), and League Two (tier 4). As is present in most soccer leagues worldwide, English soccer employs a promotion and relegation system \cite{noll2002economics}. At the end of a given season, the top teams in each league are granted promotion into the league above, while the bottom teams are sent to the league below. This provides a continuous reshuffling and rebalancing of competition.

While using a promotion and relegation system is a common feature of most soccer leagues, what distinguishes England's system from the rest of the world is revenue. The most valuable football league in the world, the Premier League, generated an impressive \euro 6.9 billion in revenue during the 22/23 season, a staggering \euro 3.1 billion more than the second highest earner, Germany’s Bundesliga \cite{Deloitte}. Promotion to the Premier League can lead to a revenue increase of at least \euro 160 million over three seasons for a newly promoted team, with that figure escalating above \euro 350 million if they are able to maintain their elevated status \cite{Deloitte}. In contrast, the lower leagues generated only \euro 890, \euro 280, and \euro 156 million respectively \cite{Deloitte}. These disparities provide strong motivation for predicting outcomes at all levels of English club soccer.

Considering the amount of quality data available describing England's lower leagues, there  has been relatively little predictive research done as of the time of writing. In 2021, Artzen and Hvattum \cite{arntzen2021predicting} employed the well-known Elo rating system to forecast match outcomes in England's lower leagues, but the application of classic linear-algebra-based models like the ones introduced by Massey \cite{massey1997} and Colley \cite{colley2002} have not been explored in the lower leagues. This research analyzes how these classical models fare in predicting end of season outcomes throughout the top four English leagues.

Given the intimate relationship of revenue with promotion and relegation, we decided to explore incorporating finances into our predictions. In 2018, Thomas Peeters used crowd-sourced team market valuations from the website Transfermarkt as a predictive factor for international soccer matches \cite{peeters2018testing}. Transfermarkt's crowd-sourced estimations come from an online forum where devoted fans debate the market value of their favorite players. Though unorthodox, the Transfermarkt values were found to have significant efficacy in predicting international soccer results, a quality Peeters’ attributes to ``wisdom of the crowd” \cite{peeters2018testing}. 

Others have explored the concept of the ``wisdom of the crowd'' in sports (e.g.  \cite{brown2019wisdom, coates2022wisdom, goldstein2014wisdom, herzog2011wisdom}) and beyond (like \cite{kremer2014implementing, lorenz2011social, mannes2014wisdom, springer2018analysis, yi2012wisdom}). For example, in \cite{herzog2011wisdom}, Herzog and Hertwig found that incorporating collective recognition into their forecasts consistently performed better than chance and similarly to predictions based on official rankings for soccer and tennis. Brown and Reade analyzed the accuracy of sports forcasts based on an online community of amateur sports tipsers called \href{https://www.oddsportal.com/}{Oddsportal}  and found that incorporating these crowd predictions can generate positive betting returns \cite{brown2019wisdom}. Supporting the findings of \cite{davis2015composition}, Brown and Reade also explored different cross sections of betters and found that incorporating the whole crowd performed better than particular subgroups even within groups of amateurs \cite{brown2019wisdom}. Finally, by analyzing Transfermarkt crowd-sourced valuations, Coates and Parshakov found that these values are correlated with but biased estimations of the actual feeds paid when a player is transferred \cite{coates2022wisdom}. Specifically, they found that the crowd-sourced metrics tend to underestimate the value of a player as indicated by Transfermarkt and this differs based on leagues \cite{coates2022wisdom}. Coates and Parshakov further suggest that Peeters' findings in \cite{peeters2018testing} could be improved by incorporating team statistics. 

In this paper, we aim to compare different models' performance in predicting different levels of English soccer. In Section \ref{sec:background}, we lay the foundation of this paper by providing an introduction to classical linear algebra models, specifically the Colley and Massey methods, as well as an overview of the key soccer player valuation site -- Transfermarkt -- we employ in this paper.
Section \ref{sec: data and evaluation} details the data and metrics used for evaluating results. In Section \ref{sec: model}, we further develop the linear models by incorporating weights and introduce the Betting Odds, Null model, and Transfermarkt regression model for subsequent comparisons.
We use weighted and unweighted Colley and Massey ranking methods to model game outcomes and end-of-season rankings. We then apply Peeters' \cite{peeters2018testing} Transfermarkt regression models to club soccer and incorporate Transfermarkt valuations into our linear algebra models.

In section \ref{sec:results}, we share the analysis of our models' predictions for English, German, and Scottish Leagues.
We conclude in Section \ref{sec:discussion} by discussing the implications of our findings on modeling different levels of soccer as well as on Transfermarkt's predictive value as an example of ``wisdom of the crowd". We then offer directions for future research. 

\section{Background}\label{sec:background}


Given the global popularity of soccer, many studies have been conducted to analyze and predict soccer team dynamics and game outcomes~\cite{arndt2016predicting, arntzen2021predicting,  harrop2014performance, hassan2020predicting, jamil2021investigation, laschober2020analysis, ramchandani2021relationship, rue2000focus, stock2022physics}. Past works have approached this problem from various directions. Ranking methods naturally arise and earlier works include the original Colley method \cite{colley2002}, Massey method \cite{massey1997, minton97}, as well as a Markov method derived from Google's PageRank algorithm \cite{langville2006google}.  Later works have continually built upon these classical models to more accurately predict not only relative ratings of teams but also individual game outcomes. For example, Govan developed the Offense-Defense Model using the Markov method to rank football teams \cite{govan2009offense}. Franceschet, Bozzo, and Vidoi introduced the temporalized version of the Massey model and applied it to the Italian Serie A soccer league \cite{bozzo2020parametric, franceschet2017temporalized}. Researchers have also employed machine learning approaches in combination with linear ranking methods to analyze soccer matches \cite{rico2023machine}. Furthermore, Kyriakides, Talattinis, and George compared linear ranking methods like Colley, Massey, and Offense-Defense to machine learning approaches like Neural Networks, Decision Trees, and Random Forests in the context of the English Premier League \cite{kyriakides2014rating}. This research aims to further develop and compare ranking methods with regression forecasting models.

\subsection{Transfermarkt} \label{sec:transfermarkt}
In this paper, we assess the predictive merit of Transfermarkt valuations. Transfermarkt is a site most notably containing user-sourced data on over 800,000 soccer players' ``market values" at a given time. The site's users deliberate on each player's value through discussion forums. No mathematical model is used to arrive at the final values; rather, a moderator evaluates users' arguments using agreed upon criteria and decides on a final player valuation (see \cite{xerxes2021} for more details). These valuations are notable for being highly regarded in the soccer community. Transfermarkt values have been cited by club executives, used by scouts, and even referenced in legal matters \cite{keppel2020}. The predictive values utilizing Transfermarkt have been a subject of interest and have been used by well-known predictive models such as the FiveThirtyEight club soccer model \cite{fivethirtyeightClubSoccer}.
Specifically, in \cite{peeters2018testing}, Thomas Peeters found predictive value comparable to betting odds and superior to traditional Elo and FIFA rankings on European and South American international matches using a model consisting of only team Transfermarkt valuations, home advantage, and the number of players on a team. 

Peeters chose to examine international soccer due to the availability of rival predictors, national teams' focus on a single competition, and the restriction in player selection mitigating endogeneity. We believe extending this analysis to club teams is important, especially in the lower leagues. 
In order to do this, we created basic models to use as rival predictors. One which we labeled the ``Null Model'' incorporated only home advantage. The other, the ``Betting Odds'' used betting odds.  More information about these models can be found in Sections \ref{sec:bettingodds} and \ref{sec: Null}.  We also implemented some classic ranking methods like the Colley and Massey method which are discussed in Sections \ref{sec: weighted colley} and \ref{sec: weighted massey}. Next, we used match-level data on lineup market value to account for differing squad selections. Furthermore, lower-level teams focus more on their league matches due to the lack of international competition and the tendency not to advance as far in inter-league domestic contests such as the FA Cup. Finally, we only aim to make predictions about future results using information available prior to a match and are not attempting make causal statements regarding market values. 

Evaluating the lower leagues using Transfermarkt valuations is particularly compelling due to the nature of the website. While player valuations are user-sourced, top players receive substantially more discussion and focus by the site's users. Lower league players often have no discussion at all and their value may solely be based on the discretion of the league's assigned moderator. For instance, the main German language market value analysis thread for the Premier League has over 900 comments at the time of writing, whereas the thread for all three lower leagues combined has only around 400, despite being up for longer.\footnote{See https://www.transfermarkt.de/england-weitere-vereine-und-themen/detail/forum/46 for an up-to-date comparison} Furthermore, since Transfermarkt aims to reflect ``adjusted medium-term demand", lower league valuations are based largely on salary and length of contract, since lower-level clubs rely mostly on free transfers rather than the large transfer fees paid at the top level.\footnote{See https://www.transfermarkt.de/marktwertanalyse-3-liga/thread/forum/67/thread\_id/237460 for an example} Therefore, market values in lower leagues are less similar to a ``wisdom of the crowd" appraisal and closer to an algorithm used by an expert. There is also a disparity in the amount of discussion between countries. The German league forums are by far the most active, while Scottish league forums generate little activity, which presents another opportunity for comparison.


\subsection{Colley Ranking Method}
\label{sec:colley}
Colley’s linear algebra-based ranking method, developed by Wesley Colley in 2002 \cite{colley2002}, is an approach to evaluate team strength. Later in Section \ref{sec: weighted colley}, we further develop and personalize this method to more accurately reflect the dynamics of the English football pyramid. The original Colley ranking method considers winning percentage and strength of schedule as the two main factors in evaluating a team's relative strength.

Often, a team's rating is tied to  win percentage so that the rating for team $i$ is calculated as $r_i=\frac{w_i}{t_i}$, where $w_i$ is the number of winning games  and $t_i$ is the total number of games played by team $i$. Colley uses Laplace's Rule of Succession to adjust win percentage, giving an untried team an initial win percentage of $50\%$. This allows him to use the following rating for team $i$: 
\begin{equation}
r_i=\frac{w_i+1}{t_i+2}. 
\end{equation}
Then, through algebraic manipulation, Colley obtains
\begin{equation}
(2+t_i)r_i=1+\frac{w_i-l_i}{2}+\frac{w_i+l_i}{2}.
\end{equation}
Now, notice that $\frac{w_i+l_i}{2}$ is simply half of the total games played by team $i$. To incorporate strength of schedule into the rating, Colley replaces this term with the sum of the ratings of the teams played by team $i$. So, he reaches the equation
\begin{equation}
    (2+t_i)r_i= 1+\frac{w_i-l_i}{2}+S
    \label{eq: colley}
\end{equation}
where $S$ is the sum of the ratings of the teams played by team $i$. This creates a symmetric system of equations,  $\mathbf{Cr = b}$, that can then be solved for the rating vector $\textbf{r}$. 


\subsection{Massey Ranking Method}\label{sec:massey}
In 1997, with an initial intent to rate college football teams, Kenneth Massey \cite{massey1997} proposed a ranking method based on score differentials. Specifically, Massey's ranking assumes that competing teams' ratings determine the point differential of their game. So for game $k$ where teams $i$ and $j$ play against each other, the margin of victory $y_k$ of the game would be 
\begin{equation}
    r_i - r_j = y_k
\end{equation}
where $r_i$ and $r_j$ represent the ratings of team $i$ and team $j$ respectively. This allows us to set up and solve the system of equations $\mathbf{Xr = y}$. Since two teams often play against each other multiple times, likely with varying point differentials, this linear system is usually inconsistent. To account for this, Massey uses least squares to approximate a solution to the system. Typically, there are infinitely many approximations that could be selected, so Massey replaces the last row with a new row which not in the span of the others in order to find a unique solution (see \cite{chartier2015life} or \cite{harsysmith2024GetInGame} for more details).  Finally, the teams are ranked by their corresponding rating in the solution vector.  

In fact, we can write the Massey system in the following form (Equation \ref{MasseyBigMatrix}) where $G_i$ denotes the total number of games played by team $i$, $g_{ij}$ denotes the number of games played between team $i$ and $j$, and $p_i$  denotes total score difference for team $i$ (sum of scores gained by team $i$ minus the sum of scores gained by its opponents).
\begin{equation}\label{MasseyBigMatrix}
\begin{bmatrix}
G_1 & -g_{12} & -g_{13} & \cdots & -g_{1n} \\
-g_{21} & G_2 & -g_{23} & \cdots & -g_{2n} \\
-g_{31} & -g_{32} & G_3 & \cdots & -g_{3n}\\
\vdots & \vdots & \vdots & \ddots & \vdots \\
-g_{n1} & -g_{n2} & -g_{n3} & \cdots & G_n
\end{bmatrix} 
\begin{bmatrix}
    r_1 \\ r_2 \\ r_3 \\ \vdots \\ r_n
\end{bmatrix} =
\begin{bmatrix}
    p_1 \\ p_2 \\ p_3 \\ \vdots \\p_n
\end{bmatrix}
\end{equation}
Given $r_i = y_k + r_j$ for game $k$, summing all games gives 
\begin{equation}
        r_i = \frac{p_i}{G_i} + \frac{1}{G_i}\sum_{j \in J} g_{ij}r_j
        \label{eq:massey rating decomp}
\end{equation}
where $J$ is the collection of teams that have competed with team $i$. Writing a team's rating as shown in Equation \ref{eq:massey rating decomp} shows that a team's rating consists of two components: the average point spread of the team and the average of its opponents' ratings.

\section{Data and Evaluation}
\label{sec: data and evaluation}


In this section, we discuss the decisions which were made pertaining to the collection of data and describe the two main methods for evaluating our models.
\subsection{Data} \label{sec:data}
We collected data on all teams from the Premier League, EFL Championship, EFL League One, and EFL League Two between 2010 and 2024. End-of-season standings were gathered from ESPN, individual match data from Football-Data.co.uk, and lineup market valuations from Transfermarkt.
In order to conduct an extended analysis, we repeated the scraping process for the top two German and top four Scottish leagues.
The lower two Scottish leagues receive little attention from Transfermarkt users; hence, most players in those leagues do not have an assigned market value. As a result, we excluded those leagues from our Transfermarkt model analysis. 
After downloading and cleaning the data, we organized it into a database for future queries.
In our dataset, there are $204$ unique teams and $47198$ games played.
See tables \ref{table:summary_all}, \ref{table:summary_eng}, \ref{table:summary_ger}, and \ref{table:summary_sco} for summary statistics by country. 

\subsection {Evaluation}
\label{sec:evaluation}
To assess the predictive power of our models, we employed two widely used measures in the field of sports analytics: ranking and game outcome predictions.

For ranking evaluation, we utilized Kendall's $\tau$ rank correlation coefficient. This metric measures the statistical association between two ordinal variables—in this case, our predicted rankings and the true rankings \cite{criado2013new, langville2012s}. The formula for Kendall's $\tau$ is given by

\begin{equation}\label{kendallT}
\tau = \frac{n_c - n_d}{n_c + n_d},
\end{equation}

\noindent where $n_c$ denotes the number of concordant pairs and $n_d$ the number of discordant pairs. Kendall's $\tau$ indicates the degree to which one ranked variable agrees with another. With values ranging between $-1$ and $1$, a higher $\tau$ indicates a stronger correlation and thus better predictive accuracy.

Additionally, we evaluated the accuracy of game outcome predictions using Brier score (see \cite{brier1950verification}).
Brier score takes the average quadratic loss for each of the three outcomes (win, loss, tie) for a given game. We use the formula in Equation \ref{eq brier score} to calculate the Brier score, $B,$ of a prediction for a specific game where $w,$ $d,$ and $l$ are $0$ or $1$ depending on whether a game ended in a win, loss, or draw. 
\begin{equation}\label{eq brier score}
B = \frac{1}{3}((p_{w} - w)^2 + (p_{d}-d)^2 + (p_{l}-l)^2)
\end{equation}  Since probabilities $p_w$, $p_d$, and $p_l$ must add up to $1$, this generates a score between $0$ and $\frac{2}{3} $, with $0$ indicating perfect prescience. A prediction of $\frac{1}{3}$ for each outcome would result in a Brier score of $0.2\bar 2$. To get the Brier score of a model across multiple predictions, one simply takes the average of its Brier scores for each individual game.

\section{Models}
\label{sec: model}
In this section, we introduce the models implemented on the obtained dataset, which will be subsequently evaluated and compared for predictive accuracy. 

\subsection{Betting Odds}\label{sec:bettingodds}
Throughout this paper, we use Betting Odds data to derive implied probabilities to which we compare our models. Our data contains decimal odds for each team winning, as well as for a draw for each game. From these odds, we can derive implied probabilities as follows: $$p_w = \frac{1}{o_w}/\Big(\frac{1}{o_w}+\frac{1}{o_d}+\frac{1}{o_l}\Big)$$ where $p_w$ is the implied probability of a win and $o_w, o_d,$ and $o_l$ are decimal odds for a win, draw, and loss respectively. The same formula can be used analogously to calculate implied probabilities of a draw and loss. In theory, given the great financial interest involved, Betting Odds can be taken as a ``best possible model,'' meaning that all information regarding a match is incorporated into odds \cite{pasteur2010extending}. While this is debatable in practice (as discussed in \cite{angelini2019efficiency}), we can use Betting Odds to measure how close our models are to this standard.

\subsection{Null Model}\label{sec: Null}
As discussed previously, we desired baseline rival models to determine whether or not our models have any substantial predictive value at all. With three outcomes, one could simply predict $\frac{1}{3}$ for a win, loss, and draw respectively. However, we can improve this by making some basic assumptions, which do not rely on specific information about any given team. For instance, we can incorporate a base rate for how often draws occur. In addition, we can use information about home advantage to slightly favor the home team according to a regression coefficient for the given league, since home advantage can vary from league to league \cite{ramchandani2021relationship}. We decided to incorporate home advantage into one of our baseline models. In order to do this, we created a model trained on previous years of each league with only information about which team played at home. For each game, we randomly assigned teams $i$ and $j,$ and use the following ordered probit regression 
\begin{equation}\label{eq:null}
   y_{ijg}^*=(h_{ig} - h_{jg})\beta_h +\epsilon_{ijg} 
\end{equation} with error term $\epsilon_{ijg}$ where $y^*_{ijg}$ is a categorical variable indicating a win, loss, or draw for team $i$ against team $j$ in game $g.$ In addition $h_{ig}$ and $h_{jg}$ are indicator variables which are $1$ if the given team ($i$ or $j$) plays game $g$ at home, and $0$ otherwise. For instance, if team $j$ plays game $g$ at home, then $h_{ig}-h_{jg}=0-1=-1.$ Finally, $\beta_h$ is an estimated coefficient for the impact of playing at home on match outcome. We expect this coefficient will be positive, which would indicate that our model thinks a home team is more likely to win. 

The ordered probit regression estimates a numerical predicted value for $y^*_{ijg}$ as well as two ``cut values." A predicted value between the two cut values indicates that the model predicts a draw, while a predicted value below or above both cut values indicates a predicted loss or win for team $i.$ More importantly, the model uses these cut values to generate predicted probabilities of a game ending each categorical outcome, which we can evaluate using Brier score as in Section \ref{sec:evaluation} (see \cite{peeters2018testing} for further discussion of ordered probit regression models as applied to soccer). 

If home teams in a given league win at different rates than visiting teams, then this model will do better than a model which predicts the same outcome for every match. We will refer to this model as the Null model, as we hope that any model which incorporates information about specific teams involved in a match should be expected to at least outperform this elementary home advantage-only model.

\subsection{Transfermarkt Regression}
\label{sec:tmreg}
We first attempt to replicate Peeters' model in \cite{peeters2018testing} with our data from English club soccer. Note that there are several important differences between our models. First, we ignore the number of players on a given team, which is less of a constraint in club soccer relative to international soccer. Second, instead of a teams' total Transfermarkt value, we obtained market value data for a team's lineup for every game in our dataset. This has the advantage of excluding players who are not on the team when a game is played or are injured for part of a season. It also ensures the most recent market value updates prior to a match are incorporated into every game's prediction. We then use the natural logarithm of a team's total lineup market value between their starters and substitutes. While using starters or players that actually played in a game may theoretically improve such a model\footnote{We did not find an improvement by doing so.}, it is not possible to know a team's starting lineup more than a few hours before a game. Thus we instead use their entire lineup, which should theoretically be more predictable on any given matchday. 

With these changes, we implemented both of Peeters' models. First, we used an ordered probit regression similar to our Null model, but with Transfermarkt values implemented as described above. The regression is
\begin{equation}
    y^*_{ijg}=(h_{ig} - h_{jg})\beta_h + (TM_{ig} - TM_{jg})\beta_{TM} + \epsilon_{ijg}.
\end{equation}

where $y_{ijg}^*, h_{ig}, h_{jg},$ and $\epsilon_{ijg}$ are defined as in our Null model (Equation \ref{eq:null}). The values of $TM_{ig}$ and $TM_{jg}$ are the natural logarithms of the total Transfermarkt value of team $i$ or team $j$'s lineup for game $g,$ and $\beta_{TM}$ is an estimated coefficient for the impact of a difference in log lineup Transfermarkt values on match outcome. Like $\beta_h,$ we expect our model to estimate a positive value of $\beta_{TM},$ which would indicate that a team with a greater lineup Transfermarkt value than their opponent is predicted to win more often.

We also implemented Peeters' ordinary least squares regression model incorporating goal difference, with an ordered probit regression performed on the predicted values from the first regression. We did not find any significant differences between the two models, so we will be using data from the first model only.

\subsection{Time-weighted Colley}
\label{sec: weighted colley}
While Colley's original method is efficient in reflecting a team's strength \cite{pasteur2010extending}, many other factors have also been shown to be significant, such as match period, match location, shots, passes, fouls, and more \cite{chartier2011NonuniWeighting, gomez2013situational, massey1997, taylor2010situation}. From multiple studies, two main weighting methods emerge in sports rating: incorporation of match time and home-field advantage \cite{chartier2011NonuniWeighting, lago2012role, langville2012s, liu2015performance, pasteur2010extending}. Weighting by match time or incorporating home advantage are natural extensions of the Colley method, so we experimented with each. Home advantage provided minimal increases in accuracy, so we restricted our weighting of the Colley method to just the date of the match. To implement this, we employ the exponential function which allows us to weigh later games more significantly than earlier games. The weight of a match $k$ is obtained from the equation 
\begin{equation}
    W_k = \exp \left( \frac{t_k-t_0}{t_f-t_0} \right)
\end{equation}
where $t_k$ is the time of match $k$, $t_0$ is the time of the earliest match, and $t_f$ is the time of the latest match in the training set. 
Now, we must incorporate these game weights into our system of linear equations. If we recall from Section \ref{sec:colley}, team $i$'s rating is equal to their adjusted win percentage, 
\begin{equation}
    r_i=\frac{w_i+1}{t_i+2}. 
\end{equation}
So, team $i$'s weighted ranking becomes 
\begin{equation}
    r^{*}_i= \frac{w^{*}_i+1}{t^{*}_i+2}
\end{equation}
where $w^{*}_i$ is the sum of the weights of the games won by team $i$ and $t^{*}_i$ is the sum of the weights of the games played by team $i$. From here, we follow a similar process as the unweighted Colley method to obtain \begin{equation}
(2+t^{*}_i)r^{*}_i=1+\frac{w^{*}_i-l^{*}_i}{2}+S^{*}
\end{equation}
where $S^{*}$ is the sum of the weighted rankings of the teams played by team $i$. Solving for the rankings follows similarly as discussed in Section \ref{sec:colley}.

Notice that this implementation of Colley's ranking method relies on the assumption of a binary match outcome, that is, one team will win and one team will lose. However, from $2010$ to $2023$, matches in the English football league ended in a draw at average rates ranging from $24\%$ in the Premier League to $27\%$ in League 1. In our model, these matches are discounted and no team is awarded merit. While this may seem as though we are discarding valuable information, omitting these matches rather than assigning an equal weighting to each team led to an increase in the predictive power of the model. Furthermore, while we experimented with assigning different weights for a drawn match based on the Transfermarkt valuation of participants, this did not seem to increase predictive power. However, this relatively high likelihood of a draw is an aspect of sports forecasting not found in many other sports and something that we hope to investigate further in the future.

\subsection{Transfermarkt-weighted Massey}\label{sec: weighted massey}
We also extended the classic Massey ranking to incorporate features tailored toward soccer matches like home advantage and match time in addition to incorporating Transfermarkt valuations 
 as discussed in Section \ref{sec:transfermarkt}.

\subsubsection{Match Time}

While constructing the linear system to obtain team ratings using the Massey Method, games from multiple years ago are weighted equally with the games played most recently. This could be problematic as more recent results may more accurately reflect a team's current strength while a game that the team played years ago might not matter at all \cite{langville2012s, pasteur2010extending, chartier2011NonuniWeighting}. Therefore, an appropriate weighting scheme should differentiate the timing of which the matches are played.  Similar to our Time-weighted Colley model, this was achieved by utilizing the exponential function which dramatically exaggerates the weight of more recent games. Specifically, we constructed the diagonal weight matrix $\mathbf{W}$ where each entry entry on the diagonal is given by
\begin{equation}
    \mathbf{W}_{kk} = \exp \left( \frac{t_k-t_0}{t_f-t_0} \right)
\end{equation}
which represents the weight applied to a specific game $k$, calculated by using the time when game $k$ is played $t_k$, the time of the first game $t_0$, and the time of the final game $t_f$. After incorporating the weight matrix, the general least-square problem transformed into a weighted least-square problem of the following form:
\begin{equation}
    \mathbf{X^\top WXr^* = X^\top W y}
\end{equation}

\noindent and solving for the match-time weighted rankings $\mathbf{r}^*$ follows similarly as discussed in Section \ref{sec:massey}.

\subsubsection{Home Advantage and Team Transfermarkt Value} 

Massey's original paper incorporated home advantage where teams are more likely to win when competing at their home stadium \cite{massey1997}. As discussed in \cite{harville1994home}, it is assumed that teams receive a fixed benefit for each home game, modifying Massey's original margin of victory equations to be
\begin{equation}
    y_k = r_i - r_j + r_hx_k,
\end{equation}
where $r_h$ is the universal home advantage and $x_k$ indicates the game location. Given the context that English league competitions usually take place at one of  the competing teams' fields, we set $x_k = 1$ if team $i$ is the home team and $x_k = -1$ otherwise.

As we did before, we can derive a new system of equations which is written below in matrix form:  
\[
    \begin{bmatrix}
    G_1 & -g_{12} & -g_{13} & \cdots & -g_{1n}  & H_1 \\
    -g_{21} & G_2 & -g_{23} & \cdots & -g_{2n} & H_2 \\
    -g_{31} & -g_{32} & G_3 & \cdots & -g_{3n} & H_3 \\
    \vdots & \vdots & \vdots & \ddots & \vdots & \vdots \\
    -g_{n1} & -g_{n2} & -g_{n3} & \cdots & G_n & H_n \\
     H_1  & H_2  & H_3 & \cdots & H_n & G_h
    \end{bmatrix} 
    \begin{bmatrix}
        r_1 \\ r_2 \\ r_3 \\ \vdots\\ r_n \\ r_h
    \end{bmatrix} =
    \begin{bmatrix}
        p_1 \\ p_2 \\ p_3 \\ \vdots \\ p_n \\ p_h
    \end{bmatrix}
\]
where each $H_i$ denotes the difference in the number of home games and away games for team $i$; $G_h$ denotes the total number of games from which a team benefits from home advantage (in our scenario, all games are played at one of two competing teams' home, so $G_h = m$ where $m$ is the total number of games); $p_h$ denotes the sum of point differentials of all games. 

We further incorporated Transfermarkt value into this model for two primary reasons. First, it enabled us to have a more comparable analysis with the Transfermarkt Regression model.  Second, after conducting regression analysis with all collected variables, results showed that average Transfermarkt values significantly predicted teams' end-of-season rankings throughout many years of data.

\begin{figure}[ht]
    \centering
    \includegraphics[width=0.9\linewidth]{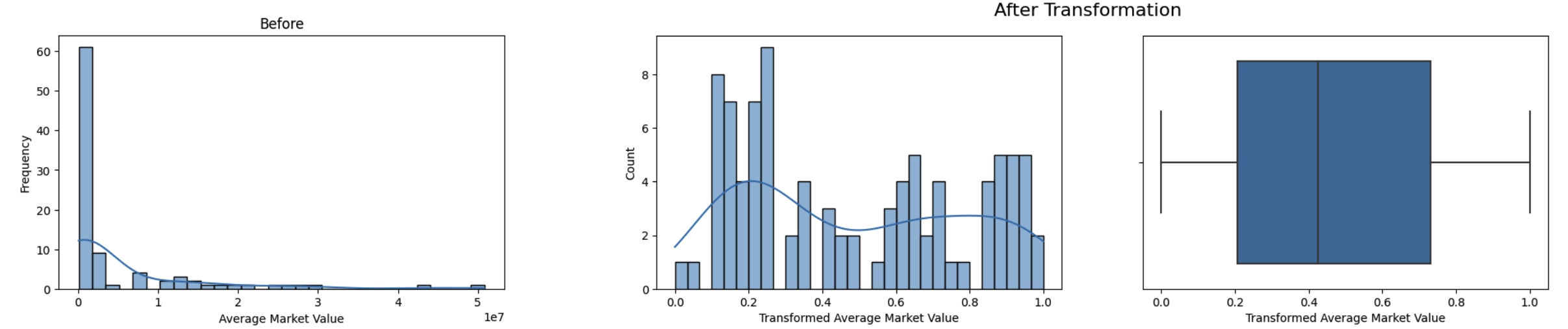}
    \caption{An example of the right-skewed average Transfermarkt value before (left) and after data transformation in 2023-24 season.}
    \label{fig:mv_distribution}
\end{figure}

Given the highly right-skewed distribution observed in Figure \ref{fig:mv_distribution}, we applied a Box-Cox transformation to normalize each year's average Transfermarkt value. This transformation was chosen after comparing its effectiveness with log, square-root, and power transformations. We further conducted standardization to ensure all teams' market value for each year ranged between 0 and 1. 

After solving the Massey system which incorporates both match time and home advantage, we obtain the solution rating $\mathbf{\hat{r}}$. Further, we combined the teams' transformed Transfermarkt values $\mathbf{\mathbf{r}}_{TM}$ with $\hat{\mathbf{r}}$ 
to obtain the final ratings $\mathbf{r}=\hat{\mathbf{r}}+\mathbf{r}_{TM}$.  
In the computation of final ratings, equal weights are assigned to the linear system solution $\mathbf{\hat{r}}$ and transformed market value $\mathbf{r}_{TM}$, since we wanted to incorporate the predicted power of the Transfermarkt site into the linear ranking methods while not making it to overly dominate the final result. 

\section{Results}
\label{sec:results}

In this section, we evaluate and compare the predictive power of our models through various metrics, including end-of-season ranking prediction and game outcome prediction, using both in-season and out-of-season data. We also discuss our results after adjusting for intra-league disparity in Section \ref{sec: adjust disparity} and extend our analysis to German and Scottish leagues in Section \ref{sec: league extension}.

\subsection{End of Season Ranking}  
\label{sec:eos-ranking}
We conducted end-of-season ranking predictions using both unweighted and weighted Colley and Massey methods and employed Kendall's $\tau$ rank correlation coefficient for accuracy evaluation. To predict the end-of-season rankings for a given year, we used all match data before the start of that season to calculate team ratings. For example, for the year 2014, we used all match data before the start of the 2014-15 season. The predicted rankings were then derived by sorting these ratings in descending order within each league. See Table~\ref{tab:raw rating data} for an example of these ratings and  Table~\ref{tab:team ranking comparison} for an example of the corresponding rankings for League One in $2023$.

\begin{table}[ht]
\small
\centering
\begin{tabular}{@{}lllll@{}}
\toprule
          & Premier League & Championship & League One & League Two \\ \midrule
Colley          & 0.5003       & 0.2076     & 0.1871   & 0.0504   \\
Massey          & 0.5498       & 0.2118     & 0.2426   & 0.1061   \\
Time-weighted Colley & 0.5004       & 0.2458     & 0.1964   & 0.0715   \\ 
T.M.-weighted Massey & 0.5887       & 0.2737      & 0.2881   & 0.1655   \\
\bottomrule
\end{tabular}
\caption{Average Kendall's $\tau$ of ranking predictions calculated from 2011-12 through 2023-24 seasons data for all leagues and models} 
\label{table:EOS}
\end{table}

Table \ref{table:EOS} summarizes the average performance of each model on each league.
Models' predictions for the Premier League significantly outperform the lower leagues, being more than twice as accurate as Championship and League One, and even more so for League Two. Within each ranking method, the weighted version achieved better accuracy than its unweighted counterpart. Furthermore, both Massey models (unweighted and Transfermarkt-weighted) performed better than the corresponding Colley models. 

\subsubsection{Game outcomes using in-season data}
\label{sec:forecast-in-season-data}
Furthermore, we used Brier scores to assess the predictability of all models on individual match outcomes. We began by training the models on the first 80\% of games within each season. Then, we used the models to predict the results of the final 20\% of games and calculated Brier scores. While ways to obtain Brier scores of forecasting methods were introduced in Section \ref{sec:evaluation}, ranking models require one extra step. After computing the ratings of each team using training data, we fitted an ordered probit regression similar to our regression in Section \ref{sec:tmreg} using the rating differentials between the home and away teams:

\begin{equation}
y^* = (r_{\text{RM}_i} - r_{\text{RM}_j})\beta_{\text{RM}}.
\end{equation}
The above regression formula is separately applied to each ranking method denoted as RM $\in $ \{Colley, Massey, Time-weighted Colley, Transfermarkt-weighted Massey\}, in which
$r_{\text{RM}_i}$ denotes the rating of home team $i$ obtained from ranking method RM, $r_{\text{RM}_j}$ denotes the rating of  away team $j$ obtained from ranking model RM, and $y^*$ is a categorical variable indicating whether the outcome is a win, draw, or loss for the home team $i$. Table \ref{table: inseason brier} shows the Brier score of each model in each English league, averaged over the 2010-11 to 2023-24 seasons.


\begin{table}[ht]
\centering
\small
\begin{tabular}{@{}lllll@{}}
\toprule
                & Premier League & Championship & League One & League Two \\ \midrule
Null            & 0.2121       & 0.2192     & 0.2167   & 0.2186   \\
Colley          & 0.1945       & 0.2156     & 0.2056    & 0.2143   \\
Massey          & 0.1912       & 0.2148     & 0.2049   & 0.2128    \\
Time-weighted Colley & 0.1946       & 0.2153     & 0.2052   & 0.2140   \\
T.M.-weighted Massey & 0.1888       & 0.2134     & 0.2040   & 0.2125   \\
T.M. Regression   & 0.1888       & 0.2145     & 0.2107   & 0.2172   \\
Betting Odds    & 0.1842       & 0.2084     & 0.2010   & 0.2067   \\ \bottomrule
\end{tabular}
\caption{Average in-season Brier score from 2010-11 to 2023-24 seasons data for all English leagues and models}
\label{table: inseason brier}
\end{table}

\subsubsection{Game outcomes using out-of season data}

Following a similar method as outlined in Section \ref{sec:forecast-in-season-data}, we further evaluated our models' predictive power on match outcomes with a larger training set. Results for our regression model, Null model, and Betting Odds remained the same. Coinciding with the results from Sections \ref{sec:eos-ranking} and \ref{sec:forecast-in-season-data}, we can see that our models performed better in the Premier League when compared to the lower three leagues.

\begin{table}[ht]
\small
\centering
\begin{tabular}{@{}lllll@{}}
\toprule
                & Premier League & Championship & League One & League Two \\ \midrule
Null            & 0.2148	       & 0.2176     & 0.2174   & 0.2186   \\                
Colley          & 0.1988       & 0.2155     & 0.2146    & 0.2188   \\
Massey          & 0.1978       & 0.2155     & 0.2144   & 0.2190    \\
Time-weighted Colley & 0.1985       & 0.2152     & 0.2145   & 0.2183   \\
T.M.-weighted Massey & 0.1955       & 0.2137     & 0.2130   & 0.2179   \\
T.M. Regression  & 0.1921       & 0.2125     & 0.2136   & 0.2173   \\
Betting Odds    & 0.1877       & 0.2081     & 0.2072   & 0.2120   \\ \bottomrule
\end{tabular}
\caption{Average Out-of-Season Brier Scores from 2012-13 to 2023-24 seasons for all English leagues and models}
\label{table: outseason brier}
\end{table}

\begin{figure}[ht]
    \centering
    \includegraphics[width=.875\linewidth]{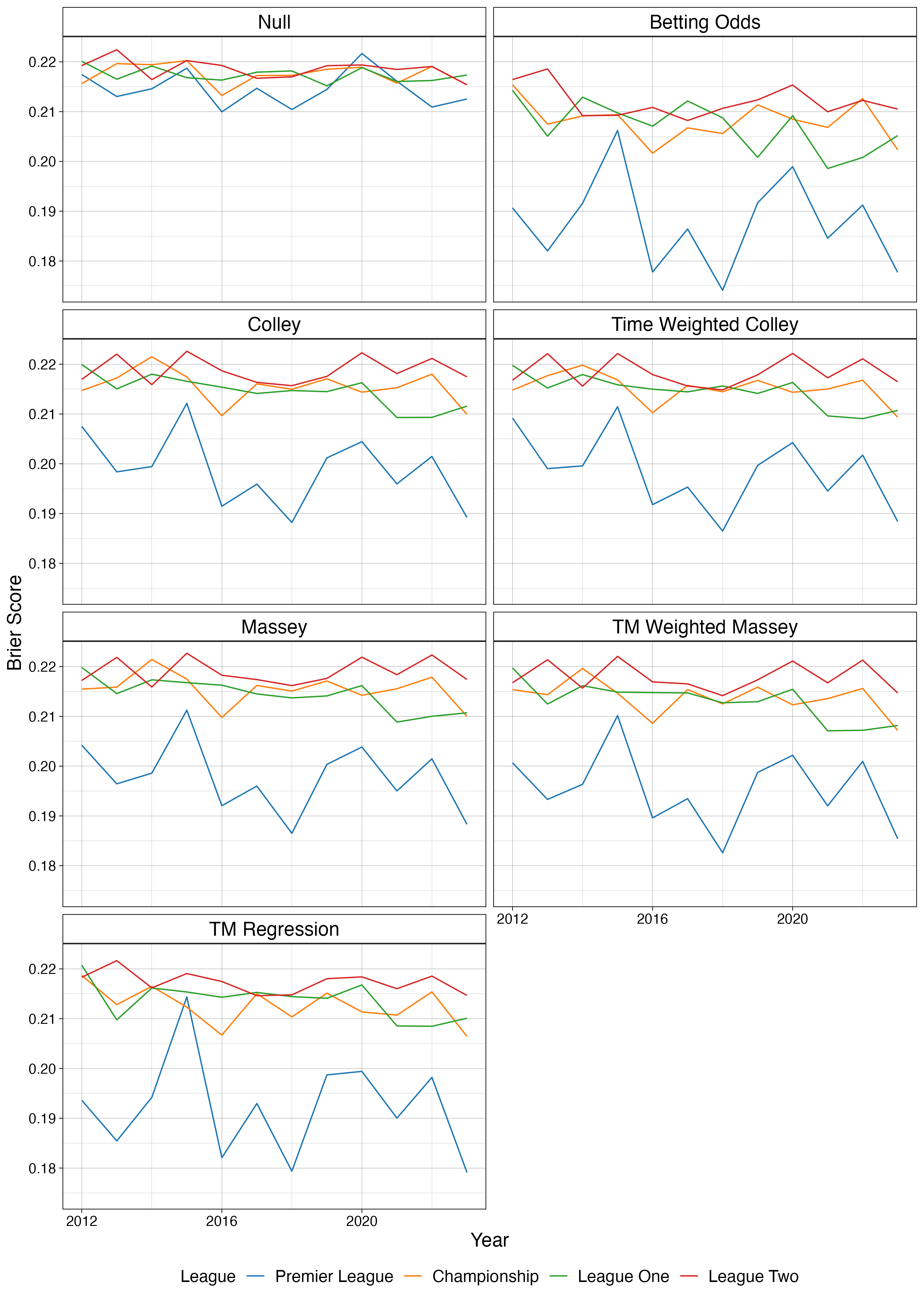}
    \caption{Brier score of out-of-season predictions using Null, Betting Odds, Colley, Time-weighted Colley, Massey, Transfermarkt-weighted Massey, Transfermarkt regression (left to right, top to bottom) ranking models from 2010-11 to 2023-24 seasons}
    \label{fig: outseason brier}
\end{figure}

After comparing all models using Brier scores computed from both in-season and out-of-season data, we found a consistent pattern. As expected, the Null model performed the worst, while the Betting Odds performed the best, with the ranking models and the Transfermarkt regression model falling in between. The Massey models (both unweighted and Transfermarkt-weighted) consistently outperformed the Colley models, with the weighted versions of both ranking models providing more accurate game outcome predictions than their unweighted counterparts. Additionally, the Transfermarkt regression model performed similarly to the Transfermarkt-weighted Massey model. To further support our observations, we conducted pairwise t-tests by game between selected models, with the results presented in Table \ref{table: out season t-test}.

\begin{table}[ht]
\centering
\small
\begin{tabular}{@{}lllll@{}}
\toprule
& \multicolumn{1}{r}{Premier League} & \multicolumn{1}{r}{Championship} & \multicolumn{1}{r}{League One} & \multicolumn{1}{r}{League Two} \\ \midrule
Colley $-$ Null                        & \textcolor{red}{-0.0159}                       & \textcolor{purple}{-0.0020}                      & \textcolor{red}{-0.0031}                   & \textcolor{blue2}{0.0003}                       \\
Massey $-$ Colley                      & \textcolor{red}{-0.0009}                       & \textcolor{blue2}{-1.6e-5}                        & \textcolor{blue2}{-0.0002}                    & \textcolor{blue}{0.0002}                     \\
Time-Colley 
$-$ Colley         & \textcolor{red}{-0.0003}                       & \textcolor{red}{-0.0004}                     & \textcolor{blue2}{-8.4e-5 }                   & \textcolor{red}{-0.0004}                   \\
T.M. Massey $-$ Massey        & \textcolor{red}{-0.0024}                    & \textcolor{red}{-0.0018}                    & \textcolor{red}{-0.0014}                  & \textcolor{red}{-0.0011}                  \\
T.M. Reg. $-$ T.M. Massey & \textcolor{blue2}{-0.0031}                          & \textcolor{blue2}{-0.0013}                        & \textcolor{blue2}{0.0008}                       & \textcolor{blue2}{-0.0007}                      \\
Betting Odds $-$ T.M. Reg.   & \textcolor{red}{-0.0044}     & \textcolor{red}{-0.0044}    & \textcolor{red}{-0.0065}   & \textcolor{red}{-0.0054}\\
Betting Odds $-$ Null                  & \textcolor{red}{-0.0268}                      & \textcolor{red}{-0.0094}                     & \textcolor{red}{-0.0103}                  & \textcolor{red}{-0.0066}                   \\ \bottomrule
\end{tabular}
\caption{Pairwise t-test comparison on Brier score between selected models, where Time-Colley stands for Time-weighted Colley, T.M. Massey stands for Transfermarkt-weighted Massey, T.M. Reg. stands for Transfermarkt Regression. Here, the first column takes the format `Model A $-$ Model B' indicating the two models being compared and the values in each cell are the Brier score difference. A negative value indicates Model A is better than Model B since a lower Brier score represents better accuracy. Following the differentials, \textcolor{red}{red} denotes $p<0.01$, \textcolor{purple}{purple} denotes $p<0.05$, \textcolor{blue}{blue} denotes $p<0.1$, \textcolor{blue2}{light blue} denotes $p\ge 0.1$.}
\label{table: out season t-test}

\end{table}

\subsection{Adjusting for intra-league disparity}
\label{sec: adjust disparity}

Such wide disparities in forecasting accuracy between the Premier League and lower leagues could have multiple explanations. First, it is possible that forecasts for the Premier League are genuinely more skilled. For instance, this could mean bookmakers incorporate more information in Premier League odds, potentially due to more data being available relative to lower leagues. For the Transfermarkt models, this could indicate that the increased attention the site's users pay to the Premier League has translated into better valuations.

A second possible explanation could be a greater disparity between teams in the Premier League. To illustrate this, consider two hypothetical matches. In the first match, suppose two identical teams play each other on neutral ground. In the second match, suppose a professional team plays against a youth team. An optimal forecast for the first match would give both teams the same chance of winning. Yet for the second match, one would predict an almost certain win for the professional team. These two forecasts would result in vastly different Brier scores, but this difference is entirely a result of the inherent disparity between the teams rather than the skill in forecasting its outcome. Directly comparing forecasting between leagues suffers from the same problem. Comparing a model to an objective baseline such as Betting Odds could remedy this, yet this just raises the question of whether Betting Odds are equally skilled across leagues, which is not obvious.

There is a strong reason to believe that this discrepancy is at least in part due to greater disparity in the Premier League. In the years for which we tested our models, the Premier League has notoriously been dominated by six clubs, often referred to in media as the ``Big Six". These clubs -- Arsenal, Chelsea, Liverpool, Manchester City, Manchester United, and Tottenham -- are wealthier, generate more revenue, and are more successful than the rest of the league (see \cite{NYTimesLeeds} and \cite{NYTimesBig6}).  From 2010 to 2024, there have only been 12 instances in which a Big Six club finished outside the top six. In contrast, teams in the lower leagues are promoted after a successful season and play against better opponents in the future, preventing similar long-term dominance from occurring.

\begin{figure}[ht]
\centering
    \centering
    \includegraphics[width=\linewidth]{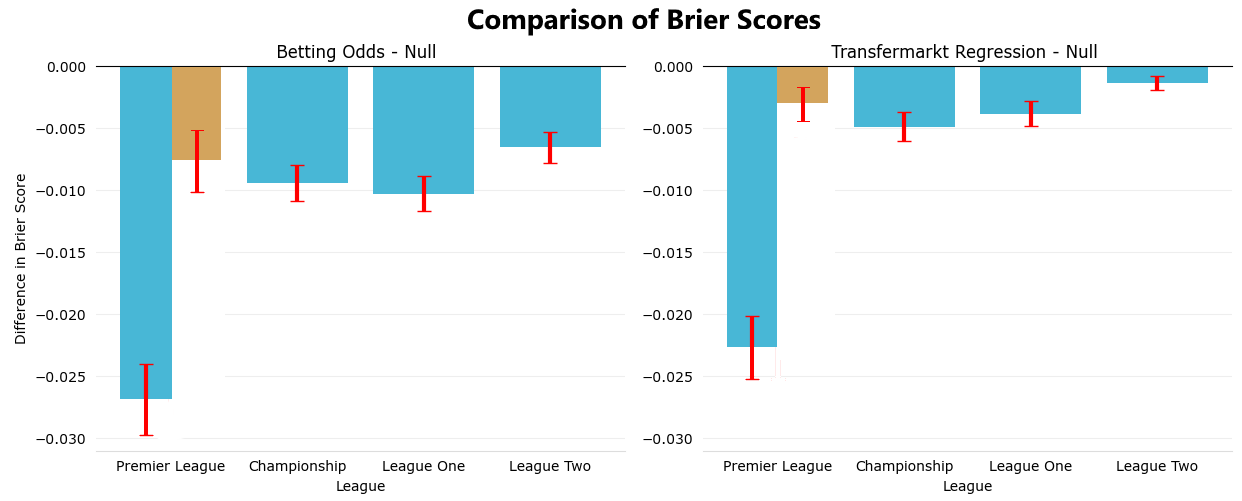}
\caption{Differences in Brier scores for Betting Odds (left) and Transfermarkt regression model (right) compared to the Null model before (blue) and after (orange) removing games involving the Big Six. Error bars indicate 95\% confidence intervals from paired t-tests.}    \label{fig:removeBigSix}
\end{figure}

To determine how much of a relative effect this disparity has, we ran our models while excluding any game involving a Big Six club. Table \ref{tab:removeBigSix} and Figure \ref{fig:removeBigSix} show the improvement in Brier score over the Null model before and after removing these games. Note that the gap between the Premier League and lower leagues is diminished.

\begin{table}[ht]
\small
\centering
\begin{tabular}{@{}lllllllll@{}}
\multicolumn{1}{c}{} & \multicolumn{2}{c}{Betting Odds} & \multicolumn{2}{c}{Transfermarkt regression} & \multicolumn{2}{c}{Weighted Colley} \\ \cmidrule(lr){2-3} \cmidrule(lr){4-5} \cmidrule(lr){6-7}
\multicolumn{1}{c|}{} & Before & \multicolumn{1}{l|}{After} & Before & After & Before & After \\ \midrule
\multicolumn{1}{l|}{Premier League} & -0.0270 & \multicolumn{1}{l|}{-0.0077} & -0.0225 & \multicolumn{1}{l|}{-0.0029} & -0.0166 & -0.0015 \\
\multicolumn{1}{l|}{Championship} & -0.0094 & \multicolumn{1}{l|}{} & -0.0049 & \multicolumn{1}{l|}{} & -0.0023 & \\
\multicolumn{1}{l|}{League One} & -0.0103 & \multicolumn{1}{l|}{} & -0.0039 & \multicolumn{1}{l|}{} & -0.0030 &  \\
\multicolumn{1}{l|}{League Two} & -0.0066 & \multicolumn{1}{l|}{} & -0.0012 & \multicolumn{1}{l|}{} & -0.0001 & \\ \bottomrule
\end{tabular}
\caption{Differences in Brier scores for Betting Odds, Transfermarkt regression, and time-weighted Colley compared to the Null model before and after removing games involving the Big Six. All differences are significant at the $p<0.001$ level.}\label{tab:removeBigSix}
\end{table}

\subsection{Extension to German and Scottish Leagues}
\label{sec: league extension}
In order to test the robustness of and extend our findings, we ran our models on data from the top two German leagues, as well as the top four Scottish leagues. We chose these two countries in order to capture specific factors not shared by English soccer. For Germany, we aimed to explore the country with the most active Transfermarkt users. Being a German site, Transfermarkt is largely dominated by German users and thus has the most discussion and interest in Germany's leagues. On the other hand, we chose Scotland as a country with a lower level of play than England\footnote{Scotland's 2023/24 UEFA coefficient ranking is 11th, compared to England in first and Germany in fourth. See https://www.uefa.com/nationalassociations/uefarankings/.} as well as less commercial interest, while also having available data on its lower leagues. In general, we found the patterns we observed among English leagues to hold across these additional countries. Table \ref{tab:GermanyScotland} shows the performance of our models on German and Scottish leagues. Note that we did not run our Transfermarkt models on Scottish League One and Two, since market valuations were rarely available for their players. Even so, we observed similar results, in that our models generally performed better in predicting the top German and Scottish leagues compared to the lower leagues.

\begin{table}[ht]
\small
\setlength{\tabcolsep}{3pt}
\centering
\resizebox{\columnwidth}{!}{
\begin{tabular}{lllllll}
\multicolumn{1}{c}{}                      & \multicolumn{2}{l}{Germany} & \multicolumn{4}{l}{Scotland} \\ \hline
\multicolumn{1}{c|}{} & Bundesliga & \multicolumn{1}{l|}{2. Bundesliga} & Premiership & Championship & League One & League Two \\ \hline
\multicolumn{1}{l|}{Null Model}           &    0.2135& \multicolumn{1}{l|}{0.2152}  &       0.2170&      0.2224&       0.2140&       0.2171\\
\multicolumn{1}{l|}{Colley}               &    0.2042& \multicolumn{1}{l|}{0.2152}  &       0.2046&      0.2169&       0.2101&       0.2225\\
\multicolumn{1}{l|}{Massey}               &    0.1991& \multicolumn{1}{l|}{0.2122}  &       0.2025&      0.2141&       0.2071&       0.2187\\
\multicolumn{1}{l|}{Time-weighted Colley} &    0.2037& \multicolumn{1}{l|}{0.2138}  &       0.2056&      0.2164&       0.2090&       0.2221\\
\multicolumn{1}{l|}{T.M.-weighted Massey} &    0.1968& \multicolumn{1}{l|}{0.2091}  &       0.2017&      0.2108& -     & -     \\
\multicolumn{1}{l|}{T.M. Regression}      &    0.1975& \multicolumn{1}{l|}{0.2089}  &       0.2003&      0.2147& -     & -     \\
\multicolumn{1}{l|}{Betting Odds}         &    0.1930& \multicolumn{1}{l|}{0.2061}  &       0.1963&      0.2048&       0.1987&       0.2092\\ \hline
\end{tabular}
}
\caption{Brier scores for our models in German and Scottish leagues. In-season models are used where applicable.}
\label{tab:GermanyScotland}
\end{table}

We further ran our models while removing top German and Scottish teams. For Scotland, this was relatively easy. As of July 2024, two clubs known collectively as the Old Firm, Celtic and Rangers, have historically dominated, making them the obvious choice when adjusting for disparities. Removing German teams was more challenging since there is no well-defined concept similar to the Big Six or Old Firm from which to draw upon. The Bundesliga has historically been dominated by Bayern Munich, winning the league in all but three years for which we have data. Two other clubs, Borussia Dortmund and Bayer Leverkusen, have also enjoyed prolonged success since 2010. For this reason, we ran our models excluding only Bayern Munich as well as excluding all three clubs. We include an aggregate standings table in the appendix (Table \ref{tab:BundesligaAGG}) to illustrate this decision. Table \ref{tab:GermanyScotlandSansTop} shows the results of this exercise, and can be compared directly to Table \ref{tab:GermanyScotland}.

\begin{table}[ht]
\small
\centering
\begin{tabular}{llll}
\multicolumn{1}{c}{}  & \multicolumn{2}{l}{Bundesliga}                             & Scottish Premiership \\ \hline
\multicolumn{1}{c|}{} & Without Bayern Munich & \multicolumn{1}{l|}{Without Top 3} & Without Old Firm     \\ \hline
\multicolumn{1}{l|}{Null Model}           &  0.2135& \multicolumn{1}{l|}{0.2148} &  0.2211\\
\multicolumn{1}{l|}{Colley}               &  0.2104& \multicolumn{1}{l|}{0.2185} &  0.2234\\
\multicolumn{1}{l|}{Massey}               &  0.2054& \multicolumn{1}{l|}{0.2140} &  0.2218\\
\multicolumn{1}{l|}{Time-weighted Colley} &  0.2099& \multicolumn{1}{l|}{0.2180} &  0.2232\\
\multicolumn{1}{l|}{T.M.-weighted Massey} &  0.2026& \multicolumn{1}{l|}{0.2117} &  0.2198\\
\multicolumn{1}{l|}{T.M. Regression}      &  0.2031& \multicolumn{1}{l|}{0.2109} &  0.2171\\
\multicolumn{1}{l|}{Betting Odds}         &  0.1991& \multicolumn{1}{l|}{0.2075} &  0.2133\\ \hline
\end{tabular}
\caption{Brier scores for our models in top German and Scottish leagues, after removing dominant teams.}
\label{tab:GermanyScotlandSansTop}
\end{table}

\section{Discussion and Conclusion}
\label{sec:discussion}

Overall, our analysis not only found disparate predictive performance of different models, it also revealed significant disparities in predictive accuracy between English leagues, as shown by multiple evaluation metrics (end-of-season ranking, in-season, and out-season Brier scores). Our models consistently demonstrated higher predictive accuracy in the Premier League compared to the lower leagues. However, this disparity diminished to insignificance when the top teams from the Premier League were excluded. Further examination of the German and Scottish Leagues revealed a similar pattern.

\subsection{Disparities between Leagues}
By exploring lower leagues, we aimed to examine what differences might show up in modeling lower-level versus elite soccer. Overall, our results indicated that any difference in model performance was due to prolonged dominance by elite teams in top leagues. When trained and tested on the entire dataset, we consistently saw our models performed better on top leagues as opposed to their lower-level counterparts. This pattern held for Betting Odds as well. However, when we only considered matches without a dominant team, there were no consistent differences in model or Betting Odds performance across leagues in any country.

Removing top teams entirely from our analysis is not without flaws. Specific seasons of lower leagues also have teams which are stronger than the rest of the league. In this sense, the models are put at a disadvantage when evaluating the Premier League, since there are fewer predictions with lower uncertainty. Still, we saw the same pattern with our Colley models, which we would not expect to perform differently in two leagues of equal parity since they are based solely on past games played between teams. In addition, directly comparing Transfermarkt models to Betting Odds without removing any teams did not indicate that Betting Odds outperformed Transfermarkt by different amounts in different leagues. These findings give us reason to believe our fundamental result holds.

\subsection{Wisdom of the Crowd}
One purpose of this paper was to apply Peeters' study of Transfermarkt valuations to club soccer. We consistently found Transfermarkt valuations to have predictive value over the Null model. However, Transfermarkt valuations were consistently outperformed by Betting Odds in club soccer, whereas the differences between Transfermarkt and Betting Odds were much closer and sometimes not statistically significant on Peeters' dataset. This could be for multiple reasons. First, our dataset was much larger, giving us more power to detect a statistically significant difference. Second, the inclusion of number of players in Peeters' model could have improved it slightly, whereas no comparable variable exists in club soccer. 

While we were able to replicate the significance of Transfermarkt valuations to club soccer, our findings regarding lower leagues cast doubt on framing Transfermarkt's predictive success as a triumph of wisdom of the crowd. If crowdsourcing player valuations led to increased predictive value, we would expect to see several results that did not show up in our analysis. First, we would expect Transfermarkt valuations in higher leagues to outperform lower leagues, given their higher amount of interest and discussion on the site. Lower league players often have no discussion at all, in which case their values are left up to the league's moderator. After removing the Big Six, we did not find any significant difference in Transfermarkt model performance between the Premier League and lower leagues. Similar adjustments in Germany and Scotland led to the same finding. 

Additionally, we would expect predictions based on Transfermarkt values in Germany and England to do better than in Scotland and potentially Germany to outperform England. Germany and England both have high levels of interest on Transfermarkt, with Germany having the most discussion. On the other hand, while Scottish leagues receive some discussion, participation in market value analysis threads for Scottish leagues and players is sparse. However, relative to Betting Odds, our Transfermarkt models do no better in German leagues than England or Scotland. 

Finally, we would expect Transfermarkt to outperform elementary models such as our Colley models, which only take into account past performance of a team. We did find this in certain leagues (English Premier League and Championship), but notably did not in either German league from our dataset, where Transfermarkt should theoretically be strongest. While some of these individual results may have other potential explanations, their combined presence indicates that crowd-sourcing is not the reason why Transfermarkt values are predictive of team success. Valuations made through discussion and deliberation by a larger number of users does not seem to be any more predictive than those made with minimal user discussion.

\subsection{Future Directions}

There are multiple ways in which we hope to extend our research. 
Our initial results show promise for incorporating market valuations as weights into linear models. While our research focused primarily on using various methods to identify trends throughout leagues, it is possible that our weighting techniques could be optimized to produce further advancements in predictive modeling of soccer matches. Likewise, adjusting the Colley method to effectively include draws could lead to improved forecasting performance. Researching relevant factors such as the implications and causes of draws throughout soccer leagues may lead to more effective models which incorporate specific characteristics of that league. Additionally, this research could be extended to explore other sports or tournaments that have a promotion and relegation system. The code for the data collection and implementation of ranking methods is available on GitHub \cite{github}.

In terms of analysis, there is room to further investigate the effect that continuously dominant teams have on the competitive balance of a league. In our models, we controlled for this by simply excluding any matches involving the ``top" teams. However, as we have discussed, this places models at a disadvantage in predicting these leagues. Thus, different approaches to controlling for disparity would provide further insight into comparing model performance between leagues.

\section{Acknowledgements}
This material is based upon work supported by the National Science Foundation under Grant No. DMS-1929284 while the authors were in residence at the Institute for Computational and Experimental Research in Mathematics in Providence, RI, during the Summer@ICERM program.

\newpage 

\bibliographystyle{plain}
\bibliography{References}

\newpage
\appendix
\section{Additional Tables}

Below are tables with the summary statistics of the leagues examined in this research. Note that in the next 4 tables, ``lineup value" is the combined market value of the starters and available substitute players for a particular game. Additionally, some variable names have been shortened. 
\begin{table}[ht]
    \centering
    \begin{tabular}{@{}llllll@{}}
        \toprule
        variable      & mean    & std     & q25     & median  & q75     \\ \midrule
        home goals    & 1.493   & 1.267   & 1       & 1       & 2       \\
        away goals    & 1.222   & 1.149   & 0       & 1       & 2       \\
        home win odds & 2.534   & 1.368   & 1.84    & 2.24    & 2.75    \\
        draw odds     & 3.658   & 0.807   & 3.29    & 3.42    & 3.68    \\
        away win odds & 3.756   & 2.523   & 2.49    & 3.12    & 4.13    \\
        lineup value  & 4.9e+07 & 1.2e+08 & 2.6e+06 & 7.2e+06 & 3.4e+07 \\
        wins          & 14.731  & 5.555   & 11      & 14      & 18      \\
        draws         & 10.189  & 3.357   & 8       & 10      & 12      \\
        losses        & 14.731  & 5.426   & 11      & 15      & 18      \\
        goals for     & 53.828  & 15.032  & 43      & 52      & 63      \\
        goals against & 53.828  & 13.580  & 45      & 54      & 63      \\
        goal diff     & 0       & 22.614  & -15     & -3      & 14      \\
        points        & 54.264  & 16.815  & 42      & 52      & 65      \\ \bottomrule
    \end{tabular}
    \caption{Summary statistics of all leagues in dataset}
    \label{table:summary_all}
\end{table}

\begin{table}[ht]
    \centering
    \begin{tabular}{@{}llllll@{}}
        \toprule
        variable      & mean    & std     & q25     & median  & q75     \\ \midrule
        home goals    & 1.447   & 1.229   & 1       & 1       & 2       \\
        away goals    & 1.179   & 1.113   & 0       & 1       & 2       \\
        home win odds & 2.479   & 1.127   & 1.87    & 2.25    & 2.74    \\
        draw odds     & 3.566   & 0.671   & 3.26    & 3.37    & 3.60    \\
        away win odds & 3.700   & 2.207   & 2.55    & 3.16    & 4.12    \\
        lineup value  & 5.7e+07 & 1.3e+08 & 2.8e+06 & 6.8e+06 & 3.8e+07 \\
        wins          & 16.185  & 5.455   & 12      & 16      & 20      \\
        draws         & 11.515  & 3.253   & 9       & 11      & 14      \\
        losses        & 16.185  & 5.416   & 12.5    & 16      & 20      \\
        goals for     & 57.611  & 14.110  & 48      & 56      & 67      \\
        goals against & 57.611  & 12.930  & 48      & 58      & 66      \\
        goal diff     & 0       & 22.288  & -16     & -2      & 15      \\
        points        & 59.926  & 16.218  & 48      & 59      & 70      \\ \bottomrule
    \end{tabular}
    \caption{Summary statistics of the top 4 English leagues}
    \label{table:summary_eng}
\end{table}

\begin{table}[ht]
    \centering
    \begin{tabular}{@{}llllll@{}}
    \toprule
        variable      & mean    & std     & q25     & median  & q75     \\ \midrule
        home goals    & 1.610   & 1.319   & 1       & 1       & 2       \\
        away goals    & 1.278   & 1.191   & 0       & 1       & 2       \\
        home win odds & 2.536   & 1.512   & 1.80    & 2.21    & 2.72    \\
        draw odds     & 3.822   & 1.018   & 3.33    & 3.51    & 3.87    \\
        away win odds & 3.950   & 2.968   & 2.53    & 3.19    & 4.32    \\
        lineup value  & 7.6e+07 & 1.2e+08 & 1.2e+07 & 2.8e+07 & 8.4e+07 \\
        wins          & 12.573  & 4.706   & 9       & 12      & 15      \\
        draws         & 8.853   & 2.636   & 7       & 9       & 11      \\
        losses        & 12.573  & 4.366   & 10      & 13      & 15      \\
        goals for     & 49.089  & 14.207  & 39      & 47      & 56      \\
        goals against & 49.089  & 11.080  & 42      & 49      & 56      \\
        goal diff     & 0       & 21.438  & -14     & -3      & 12.25   \\
        points        & 46.548  & 13.294  & 37      & 44      & 55      \\ \bottomrule
    \end{tabular}
    \caption{Summary statistics of the top 2 German leagues}
    \label{table:summary_ger}
\end{table}

\begin{table}[ht]
    \centering
    \begin{tabular}{@{}llllll@{}}
        \toprule
        variable      & mean    & std     & q25     & median  & q75     \\ \midrule
        home goals    & 1.524   & 1.317   & 1       & 1       & 2       \\
        away goals    & 1.292   & 1.204   & 0       & 1       & 2       \\
        home win odds & 2.678   & 1.772   & 1.80    & 2.24    & 2.83    \\
        draw odds     & 3.775   & 0.906   & 3.34    & 3.47    & 3.78    \\
        away win odds & 3.745   & 2.895   & 2.31    & 2.95    & 4.04    \\
        lineup value  & 4.9e+06 & 1.2e+07 & 2.2e+05 & 2.2e+06 & 4.9e+06 \\
        wins          & 13.396  & 5.515   & 10      & 12      & 16      \\
        draws         & 8.432   & 2.835   & 6       & 8       & 10      \\
        losses        & 13.396  & 5.314   & 10      & 14      & 16      \\
        goals for     & 49.607  & 15.418  & 39      & 47.5    & 58      \\
        goals against & 49.607  & 14.393  & 39      & 50      & 59      \\
        goal diff     & 0       & 24.279  & -15     & -3      & 12      \\
        points        & 48.486  & 16.216  & 38      & 46      & 58      \\ \bottomrule
    \end{tabular}
    \caption{Summary statistics of the top 4 Scottish leagues}
    \label{table:summary_sco}
\end{table}

\begin{table}[ht]
\centering
\begin{tabular}{@{}llllllll@{}}
\toprule
\# & Club                     & Matches & W   & D   & L   & GD   & Points \\ \midrule
1  & Bayern Munich            & 476     & 347 & 69  & 60  & 845  & 1,110  \\
2  & Borussia Dortmund        & 476     & 285 & 93  & 98  & 489  & 948    \\
3  & Bayer 04 Leverkusen      & 476     & 246 & 104 & 126 & 280  & 842    \\
4  & Borussia Mönchengladbach & 476     & 195 & 117 & 164 & 81   & 702    \\
5  & VfL Wolfsburg            & 476     & 179 & 125 & 172 & 13   & 662    \\
6  & TSG 1899 Hoffenheim      & 476     & 167 & 130 & 179 & 0    & 631    \\
7  & 1.FSV Mainz 05           & 476     & 160 & 121 & 195 & -75  & 601    \\
8  & Eintracht Frankfurt      & 442     & 155 & 124 & 163 & -33  & 589    \\
9  & SC Freiburg              & 442     & 151 & 121 & 170 & -99  & 574    \\
10 & FC Schalke 04            & 408     & 150 & 98  & 160 & -58  & 548    \\
11 & SV Werder Bremen         & 442     & 133 & 118 & 191 & -157 & 517    \\
12 & RB Leipzig               & 272     & 147 & 65  & 60  & 233  & 506    \\
13 & FC Augsburg              & 442     & 130 & 114 & 198 & -164 & 504    \\
14 & VfB Stuttgart            & 408     & 136 & 94  & 178 & -84  & 502    \\
15 & Hertha BSC               & 374     & 112 & 94  & 168 & -148 & 430    \\
16 & 1.FC Köln                & 374     & 104 & 106 & 164 & -165 & 418    \\
17 & Hannover 96              & 272     & 87  & 56  & 129 & -112 & 317    \\
18 & Hamburger SV             & 272     & 79  & 64  & 129 & -146 & 301    \\ \bottomrule
\end{tabular}
\caption{Aggregate standings for the Bundesliga from 2010 through 2024. Only the top 18 teams for this period are shown. Data compiled using Transfermarkt's eternal table tool.}
\label{tab:BundesligaAGG}
\end{table}

\begin{table}[]
\begin{tabular}{lll}
\hline
League One Team     & TM-weighted Massey Rating & Time-weighted Colley Rating \\ \hline
Barnsley            & 0.465751477               & 10.81142992                 \\
Blackpool           & 0.468738141               & 10.78199041                 \\
Bolton Wanderers    & 0.371369091               & 10.78576544                 \\
Bristol Rovers      & -0.104628922              & 10.66566011                 \\
Burton Albion       & -0.073532264              & 10.72761681                 \\
Cambridge United    & -0.245390316              & 10.63263385                 \\
Carlisle United     & -0.287612208              & 10.6719491                  \\
Charlton Athletic   & 0.306562569               & 10.80338532                 \\
Cheltenham Town     & -0.186331949              & 10.66652252                 \\
Derby County        & 0.862484119               & 10.9013147                  \\
Exeter City         & 0.019166283               & 10.70872262                 \\
Fleetwood Town      & 0.04559182                & 10.71850666                 \\
Leyton Orient       & -0.121947198              & 10.66637537                 \\
Lincoln City        & 0.142115261               & 10.70848642                 \\
Northampton Town    & -0.165362375              & 10.68197785                 \\
Oxford United       & 0.217674326               & 10.72990384                 \\
Peterborough United & 0.48397419                & 10.76550677                 \\
Port Vale           & -0.168985816              & 10.65510619                 \\
Portsmouth          & 0.410939337               & 10.78575663                 \\
Reading             & 0.700048207               & 10.85541182                 \\
Shrewsbury Town     & -0.041858943              & 10.70904011                 \\
Stevenage           & -0.255176713              & 10.64926415                 \\
Wigan Athletic      & 0.582918513               & 10.83900958                 \\
Wycombe Wanderers   & 0.083994503               & 10.74005287                 \\ \hline
\end{tabular}
\caption{League One predictive ratings for $2023$ calculated using the end of season Transfermarkt-weighted Massey and Time-weighted Colley methods (as  described at the start of  Section~\ref{sec:eos-ranking}).}
\label{tab:raw rating data}
\end{table}

\begin{table}[]
\begin{tabular}{llll}
\hline
Ranking & League One Official & T.M.-weighted Massey & Time-weighted Colley \\ \hline
1       & Portsmouth          & Derby County         & Derby County         \\
2       & Derby County        & Reading              & Reading              \\
3       & Bolton Wanderers    & Wigan Athletic       & Wigan Athletic       \\
4       & Peterborough United & Peterborough United  & Barnsley             \\
5       & Oxford United       & Blackpool            & Charlton Athletic    \\
6       & Barnsley            & Barnsley             & Bolton Wanderers     \\
7       & Lincoln City        & Portsmouth           & Portsmouth           \\
8       & Blackpool           & Bolton Wanderers     & Blackpool            \\
9       & Stevenage           & Charlton Athletic    & Peterborough United  \\
10      & Wycombe Wanderers   & Oxford United        & Wycombe Wanderers    \\
11      & Leyton Orient       & Lincoln City         & Oxford United        \\
12      & Wigan Athletic      & Wycombe Wanderers    & Burton Albion        \\
13      & Exeter City         & Fleetwood Town       & Fleetwood Town       \\
14      & Northampton Town    & Exeter City          & Shrewsbury Town      \\
15      & Bristol Rovers      & Shrewsbury Town      & Exeter City          \\
16      & Charlton Athletic   & Burton Albion        & Lincoln City         \\
17      & Reading             & Bristol Rovers       & Northampton Town     \\
18      & Cambridge United    & Leyton Orient        & Carlisle United      \\
19      & Shrewsbury Town     & Northampton Town     & Cheltenham Town      \\
20      & Burton Albion       & Port Vale            & Leyton Orient        \\
21      & Cheltenham Town     & Cheltenham Town      & Bristol Rovers       \\
22      & Fleetwood Town      & Cambridge United     & Port Vale            \\
23      & Port Vale           & Stevenage            & Stevenage            \\
24      & Carlisle United     & Carlisle United      & Cambridge United     \\ \hline
\end{tabular}
\caption{League One predictive rankings for $2023$ calculated using the end of season Transfermarkt-weighted Massey and Time-weighted Colley methods (as  described at the start of  Section~\ref{sec:eos-ranking}) compared to the EFL Official rankings.}\label{tab:team ranking comparison}
\end{table}

\end{document}